\documentclass[%
 reprint,
superscriptaddress,
 amsmath,amssymb,
 aps,
]{revtex4-1}

\usepackage{graphicx}% Include figure files
\usepackage{dcolumn}% Align table columns on decimal point
\usepackage{bm}% bold math
\newcommand{\pd}{\partial}

\newcommand{\pvec}{{\boldsymbol{p}}}

\newcommand{\pb}{{\bar{p}}}

\newcommand{\PP}{{\mathbb{P}}}

\newcommand{\PPb}{{\overline{\mathbb{P}}}}

\newcommand{\perm}{{\text{perm}}}

\newcommand{\hsgra}{\text{HiSGRA}}

\begin{document}

%\preprint{LMU-ASC ??/18}

\title{Quantum Chiral Higher Spin Gravity}
\author{Evgeny Skvortsov}
\affiliation{%
Albert Einstein Institute,\\
Am M\"{u}hlenberg 1, D-14476, Potsdam-Golm, Germany 
}%
\affiliation{Lebedev Institute of Physics, \\
Leninsky ave. 53, 119991 Moscow, Russia}
\email{evgeny.skvortsov@aei.mpg.de}
\author{Tung Tran}
\affiliation{%
 Arnold Sommerfeld Center for Theoretical Physics,\\
 Ludwig Maximilian University of Munich,\\
Theresienstr.  37, D-80333 M\"unchen, Germany
}%
\email{tung.tran@lmu.de}
\author{Mirian Tsulaia}
\affiliation{
School of Physics M013, The University of Western Australia, 35 Stirling
Highway, Crawley, Perth, WA 6009, Australia}
\email{mirian.tsulaia@gmail.com}

\date{\today}% It is always \today, today,

\begin{abstract}
An example of a higher spin gravity in four-dimensional flat space has recently been constructed in \cite{Ponomarev:2016lrm}. This theory is chiral and the action is written in the light-cone gauge. The theory has certain stringy features, e.g. admits Chan-Paton factors. We show that the theory is consistent, both at the classical and quantum level. Even though the interactions are non-trivial, due to the coupling conspiracy all tree level amplitudes vanish on-shell. The loop corrections also vanish. Therefore, the full quantum S-matrix is one and the theory is consistent with the numerous no-go theorems. This provides the first example of a (quantum) interacting higher spin gravity with an action. We argue that higher spin gravities in AdS space should display the same features.

\pacs{04.62.+v,\, 11.25.Hf,\, 11.25.Tq}% PACS, the Physics and Astronomy
\keywords{QFT, Quantum Gravity, Higher Spin Gravity}
\end{abstract}

\maketitle

%\tableofcontents
%%%%%%%%%%%%%%%%%%%%%%%%%%%%%%%%%%%%%%%%%%%%%%%%%%%%%%%%%%
\section{\label{sec:intro}Introduction}
%%%%%%%%%%%%%%%%%%%%%%%%%%%%%%%%%%%%%%%%%%%%%%%%%%%%%%%%%%
Higher spin gravities (\hsgra) are hypothetical theories that contain graviton and massless fields with spin greater than two. \hsgra's have a checkered past, since they had long been believed not to exist due to many no-go theorems \footnote{Incomplete list includes \cite{Weinberg:1964ew,Coleman:1967ad,Bekaert:2010hp} and review \cite{Bekaert:2010hw}.}: most notably the Weinberg low energy theorem \cite{Weinberg:1964ew} and the Coleman-Mandula theorem \cite{Coleman:1967ad}. The theorems directly constrain the $S$-matrix: the former does not allow massless fields with spin greater than two to couple nontrivially to the usual low spin particles, and the latter prevents the S-matrix from having symmetry generators that transform as tensors under the Lorentz group. 

While the theorems restrict the footprints of interactions at infinity, they have little to say about local effects. Intriguingly, massless higher spin fields were shown to have some consistent local cubic interactions, with the first positive results having been obtained in the light-cone gauge \cite{Bengtsson:1983pg,Bengtsson:1983pd}. One of the greatest advantages of the light-cone approach is that it operates only with physical degrees of freedom with the idea of explicitly constructing field dependent realization of the Poincare algebra.

However, the existence of cubic vertices does not yet guarantee that the higher point amplitudes respect Poincare symmetries. The analysis of closure of the Poincare algebra at the quartic order was performed in \cite{Metsaev:1991nb,Metsaev:1991mt} and left some possibilities open. Recently in \cite{Ponomarev:2016lrm}, which is heavily based on \cite{Metsaev:1991nb,Metsaev:1991mt}, it was shown that there exists a simple solution for the Poincare algebra generators to all orders. The solution was called chiral \hsgra{} since its interaction vertices discriminate between helicities: there are more fields with positive helicities than with negative. 

Despite its simplicity, the chiral \hsgra{} has all the features that \hsgra's are expected to have on general grounds. The spectrum consists of all massless integer spin fields, starting with the scalar field. The graviton is a part of this spectrum and the theory has a dimensionful coupling constant, which can be identified with the Planck length, $l_p$. The theory admits Yang-Mills gaugings \footnote{$o(N)$ color factors were introduced in \cite{Metsaev:1991mt}, but it is easy to see that $u(N)$ and $usp(N)$ work as well. }, which follow the stringy Chan-Paton pattern. All fields have nontrivial (self-  and gravitational) interactions that are crucial for classical consistency.

The purpose of this letter is to report that the chiral \hsgra{} is consistent both at the classical and quantum levels. Moreover, it is not in contradiction with the no-go theorems: the couplings conspire in such a way that the full S-matrix is $1$ and therefore, when observed at infinity, higher spin fields appear to have trivial scattering. Nevertheless, the chiral \hsgra{} provides the first example of a quantum \hsgra. Moreover, it is the first example of an interacting \hsgra{} whose action can explicitly be written. We will also argue that the AdS \hsgra{} counterparts should follow the same pattern, though in a more complicated way.

%%%%%%%%%%%%%%%%%%%%%%%%%%%%%%%%%%%%%%%%%%%%%%%%%%%%%%%%%%
\section{\label{sec:hsgra}Chiral Higher Spin Gravity}
%%%%%%%%%%%%%%%%%%%%%%%%%%%%%%%%%%%%%%%%%%%%%%%%%%%%%%%%%%
Chiral \hsgra{} is known in the light-cone gauge and we  present the action directly in momentum space to facilitate the computation of amplitudes. The $4d$ momentum is $\pvec=(p^+,p^-,p,\bar{p})$ and $p^+$ is usually denoted by $\beta$. In the light-cone gauge a massless spin-$s$ field is represented by a pair of scalar fields: $\Phi^{\pm s}_\pvec\equiv \Phi^{\pm s}(\pvec)$, $\Phi^{+s}(\pvec)^\dag=\Phi^{-s}(\pvec)$. 
It is also possible and computationally convenient to consider a version of the theory, the 'higher spin glue', by taking $\Phi_s$ to be $u(N)$-valued. The action of the chiral \hsgra{} in momentum space reads:
\begin{widetext}
\begin{align}
\begin{aligned}
S&=-\sum_{\lambda}\int d^4 \pvec\, \mathrm{Tr}[\Phi^{\lambda}_{\pvec}{}^\dag \Phi^\lambda_{\pvec}]\,\pvec^2 +\sum_{\lambda_{1,2,3}}C_{\lambda_1,\lambda_2,\lambda_3}\int d^4 \pvec_{1,2,3}\,  \frac{\PPb^{\lambda_1+\lambda_2+\lambda_3}}{\beta_1^{\lambda_1}\beta_2^{\lambda_2}\beta_3^{\lambda_3}}\mathrm{Tr}[\Phi^{\lambda_1}_{\pvec_1}\Phi^{\lambda_2}_{\pvec_2}\Phi^{\lambda_3}_{\pvec_3}]\,\delta^{4}(\pvec_1+\pvec_2+\pvec_3)\,,
\end{aligned}
\end{align}
\end{widetext}
where $\mathrm{Tr}$ is the trace over the implicit $u(N)$ indices, $\PPb\equiv\PPb_{12}\equiv \pb_1 \beta_2-\pb_2\beta_1$. It is crucial for the closure of the Poincare algebra to choose the coupling constants as \cite{Metsaev:1991nb,Metsaev:1991mt,Ponomarev:2016lrm}
\begin{align}\label{couplings}
    C_{\lambda_1,\lambda_2,\lambda_3}&=\frac{(l_p)^{\lambda_1+\lambda_2+\lambda_3-1}}{\Gamma(\lambda_1+\lambda_2+\lambda_3)}\,.
\end{align}
The $\Gamma$-factor requires the sum over helicities in the vertex to be positive and triplets $+++$, $++-$ and $+--$ are present in general. There is a dimensionful coupling constant, $l_p$, to be associated with the Planck length as the $(+2,+2,-2)$ vertex present here is a part of the usual Einstein-Hilbert action. The light-cone approach is very close to the spinor-helicity formalism, as it was noted e.g. in \cite{Chakrabarti:2005ny,Ananth:2012un,Bengtsson:2016jfk,Ponomarev:2016cwi}, and the vertex has a clear interpretation
\begin{align}\notag
\frac{\PPb^{\lambda_1+\lambda_2+\lambda_3}}{\beta_1^{\lambda_1}\beta_2^{\lambda_2}\beta_3^{\lambda_3}} \sim 
    [12]^{\lambda_1+\lambda_2-\lambda_3}[23]^{\lambda_2+\lambda_3-\lambda_1}[13]^{\lambda_1+\lambda_3-\lambda_2}\,.
\end{align}
Lastly, presence of the Chan-Paton factors leads to significant simplifications as only color-ordered amplitudes need to be computed.

%%%%%%%%%%%%%%%%%%%%%%%%%%%%%%%%%%%%%%%%%%%%%%%%%%%%%%%%%%
\section{\label{sec:tree}Tree Amplitudes}
%%%%%%%%%%%%%%%%%%%%%%%%%%%%%%%%%%%%%%%%%%%%%%%%%%%%%%%%%%
We would like to compute all physical $n$-point tree amplitudes, i.e. amplitudes $A_n(\pvec_1,...,\pvec_n)$ with external legs being on-shell, $\pvec_i^2=0$. It turns out that all $n$-point amplitudes can be computed recursively if lower order amplitudes with one external off-shell leg are known:
\begin{align}\label{boots}
    \parbox{4.8cm}{\includegraphics[scale=0.2]{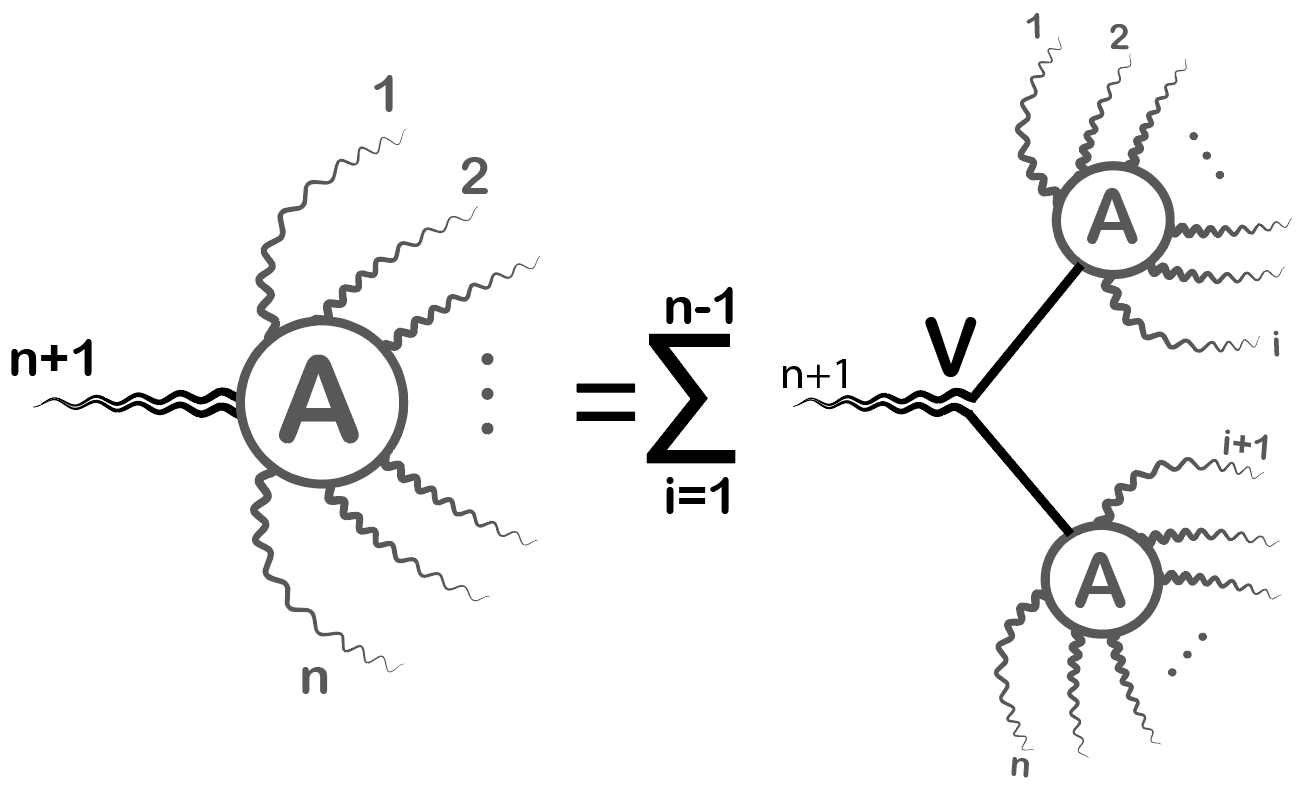}}
\end{align}
The picture illustrates that the $(n+1)$-point amplitude with one off-shell leg can be obtained as a sum over all ways to attach $(i+1)$- and $(n+1-i)$-point amplitudes to the cubic vertex. The legs being attached have to be off-shell, which explains why we need to know lower order amplitudes with just one off-shell leg. The simplest amplitude is the $4$-point one \footnote{See also \cite{Ponomarev:2016lrm} for the uncolored case.}
\begin{align}\notag
    \parbox{1.5cm}{\includegraphics[scale=0.13]{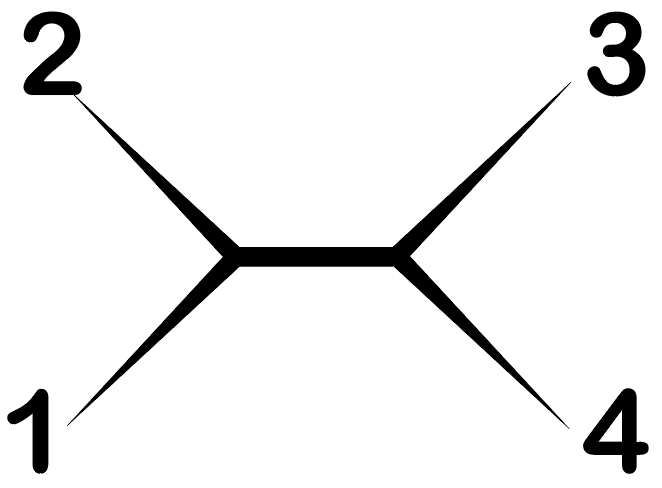}}+\parbox{1.4cm}{\includegraphics[scale=0.13]{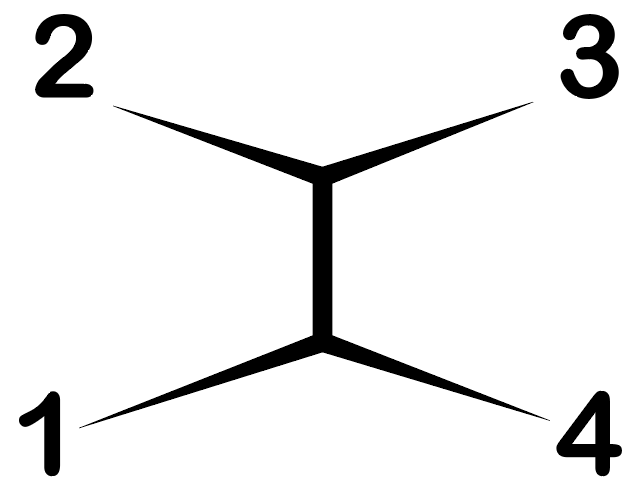}}=\mathcal{N}_4\frac{\alpha_4^{\Lambda_4-2}\beta_2\, \pvec_4^2}{\beta_4\PP_{12}\PP_{23}}\,,
\end{align}
where we define $\Lambda_n=\lambda_1+...+\lambda_n$,
\begin{align}
    \mathcal{N}_n&=\frac{(-)^n}{2^{n-2}\Gamma(\Lambda_n-(n-3))\prod_{i=1}^{n}\beta_i^{\lambda_i-1}}
\end{align}
and $\alpha_n=\sum_{i<j}^{n-2}\PPb_{ij}+\PPb_{n-1,n}$, e.g. $\alpha_4=\PPb_{12}+\PPb_{34}$.

Thanks to the $\pvec_4^2$ factor, the physical amplitude $A_4$ vanishes. Now it is a matter of direct computation to prove, with the use of \eqref{boots}, that the $n$-point amplitude is 
\begin{align}\notag
    A_n=\mathcal{N}_n\frac{\alpha_n^{\Lambda_n-(n-2)} \beta_2...\beta_{n-2}\,\pvec_n^2}{\beta_n\PP_{12}...\PP_{n-2,n-1}}\,.
\end{align}
Again, we see that it has $\pvec_n^2$ factor and therefore the physical $n$-point amplitude $A_n$ vanishes. This makes the chiral \hsgra{} consistent with the no-go theorems at least at the tree level. It is worth stressing that such a simple result for amplitudes with one off-shell leg relies on the particular form of the coupling constants in \eqref{couplings}. We call this situation coupling conspiracy, since the multitude of nontrivial interactions conspire to cancel in the physical answers. Lastly, vanishing of tree level amplitudes should improve the UV behaviour of loop diagrams. 

%%%%%%%%%%%%%%%%%%%%%%%%%%%%%%%%%%%%%%%%%%%%%%%%%%%%%%%%%%
\section{\label{sec:loops}Vacuum Loops}
%%%%%%%%%%%%%%%%%%%%%%%%%%%%%%%%%%%%%%%%%%%%%%%%%%%%%%%%%%
Vacuum diagrams play an important role in the cancellation of legged loop diagrams in the chiral \hsgra. If we had a covariant action for the theory \footnote{In the covariant description a massless spin-$s$ field is usually represented by a rank-$s$ symmetric tensor and has a gauge symmetry with a rank-$(s-1)$ parameter \cite{Fronsdal:1978rb}.}, the one-loop vacuum bubble would be equal to the product of determinants of the kinetic terms \footnote{Dependence on $N$ is not important for what follows.}
\begin{align}\notag
    \parbox{0.8cm}{\includegraphics[scale=0.15]{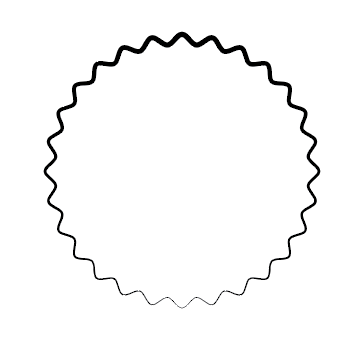}}&:  &
   Z_{\text{1-loop}}&= \frac{1}{(z_0)^{1/2}}\prod_{s>0}\frac{(z_{s-1})^{1/2}}{(z_s)^{1/2}}\,,
\end{align}
where $z_s=\det_{s,\perp}|-\pd^2|$ is the determinant of $-\pd^2$ on the space of transverse and traceless tank-$s$ tensors. The numerator results from the kinetic terms of ghosts. Such determinants have been already studied, both in flat and AdS spaces \footnote{For one-loop tests of \hsgra, see e.g. \cite{Giombi:2013fka,Giombi:2016pvg,Giombi:2014yra,Gunaydin:2016amv,Bae:2016rgm,Skvortsov:2017ldz}}. The lesson is that the sum (product) over spins needs to be regularized. In our case it is possible \footnote{There are many nontrivial one-loop examples that make us confident that the seemingly ad-hoc regularization below is correct, see \cite{Beccaria:2015vaa}.} to adopt a regularization that makes the cancellation between all numerators and denominators obvious. The same partition function can be interpreted as $Z_{\text{1-loop}}=(z_0)^{\nu_0}$, $\nu_0$ being the total number of degrees of freedom, which can be regularized as
\begin{align}\label{magic}
    &\nu_0=\sum_\lambda 1=1+2 \sum_{s=1}^{\infty}1=1+2\zeta(0)=0\,,
\end{align}
i.e. we interpret the spectrum as a scalar field plus $2$ degrees of freedom per each of $s>0$ field. Then, we find $Z_{\text{1-loop}}=(z_0)^{\nu_0}=1$ \cite{Beccaria:2015vaa}. 

All the other vacuum diagrams vanish without a need to regularize. Indeed, the sum of helicities over all the vertices has to be zero, while for a vertex to contribute the sum over the three ingoing helicities must be positive due to the $\Gamma$-factor in \eqref{couplings}. Therefore, there is always at least one vertex where the $\Gamma$-factor makes the whole diagram vanish. 

To summarize, all vacuum diagrams vanish: the one-loop diagram after the appropriate regularization and all the others due to the coupling conspiracy.

%%%%%%%%%%%%%%%%%%%%%%%%%%%%%%%%%%%%%%%%%%%%%%%%%%%%%%%%%%
\section{\label{sec:loops}Legged Loops}
%%%%%%%%%%%%%%%%%%%%%%%%%%%%%%%%%%%%%%%%%%%%%%%%%%%%%%%%%%
We would like to examine the behaviour of legged loop diagrams and see if the coupling conspiracy makes them vanish in one way or another. Two lower order amplitudes are considered in detail and then the general argument is given. We expect the loops not to have any cuts due to the vanishing of tree level amplitudes.

%%%%%%%%%%%%%%%%%%%%%%%%%%%%%%%%%%%%%%%%%%%%%%%%%%%%%%%%%%
\subsection{Self Energy}
%%%%%%%%%%%%%%%%%%%%%%%%%%%%%%%%%%%%%%%%%%%%%%%%%%%%%%%%%%
The self-energy diagram, the bubble, is the one we might expect to be UV divergent
\begin{align}\notag
    \parbox{1.9cm}{\includegraphics[scale=0.115]{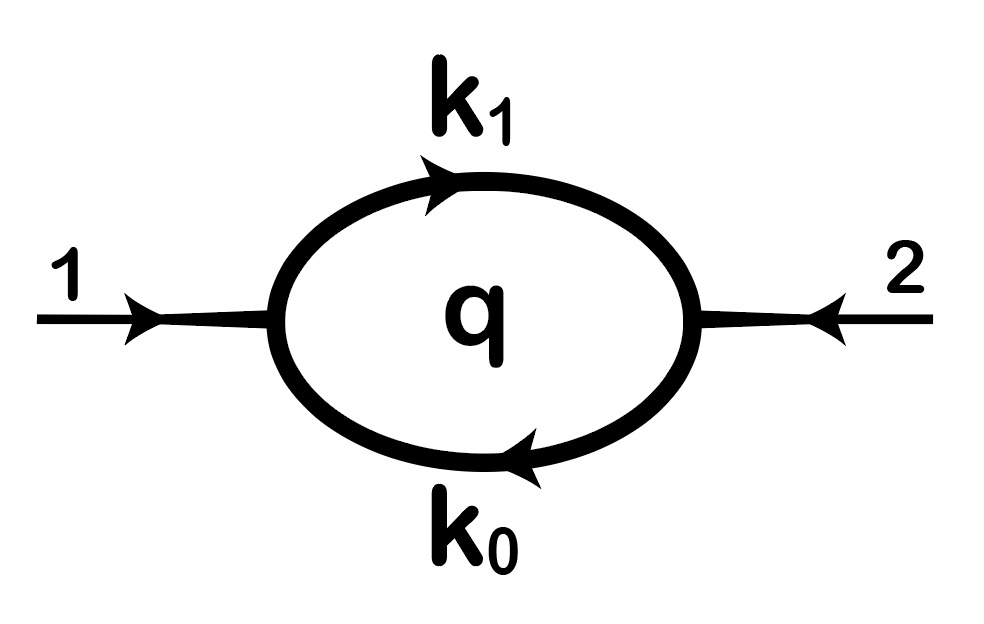}}=\frac{\nu_0(l_p)^{\Lambda_2-2}}{\Gamma[\Lambda_2-1]}\int\frac{d^4 q}{(2\pi)^4} \frac{\PPb_{k_0-q,p}^{2}\delta_{\Lambda_2,2}}{(q-k_0)^2(q-k_1)^2}\,,
\end{align}
where we dropped few unessential factors and $k_{0,1}$ are the dual momenta, $\pvec=k_1-k_0$. We prefer to use the worldsheet-friendly regularization \cite{Thorn:2004ie,Chakrabarti:2005ny}, which was shown to work nicely in a number of theories, including QCD. The main feature is that $\nu_0$ factors out, which allows us to declare vanishing of the whole diagram. Let us, nevertheless, evaluate the integral. For the physical amplitude $\pvec^2=0$ the result is simple and finite
\begin{align}
  \int_0^1 dx [x \bar{k}_0+(1-x)\bar{k}_1]^{2}\,,
\end{align}
which is reminiscent of the $\Pi^{++}$ amplitude in \cite{Chakrabarti:2005ny} for $\Lambda_2=2$. If it were not for $\nu_0=0$ we would have to add a counterterm to eliminate the correction above since it breaks Lorentz invariance. 

%%%%%%%%%%%%%%%%%%%%%%%%%%%%%%%%%%%%%%%%%%%%%%%%%%%%%%%%%%
\subsection{Vertex Correction}
%%%%%%%%%%%%%%%%%%%%%%%%%%%%%%%%%%%%%%%%%%%%%%%%%%%%%%%%%%
The physical three-point amplitude for massless spinning fields is zero for kinematical reasons and we keep one momentum off-shell. In the large-$N$ \footnote{The large-$N$ limit does not affect our general argument that the sum over helicities will factor out, see \cite{STT}.} the result is
\begin{align}\notag
    \parbox{1.9cm}{\includegraphics[scale=0.116]{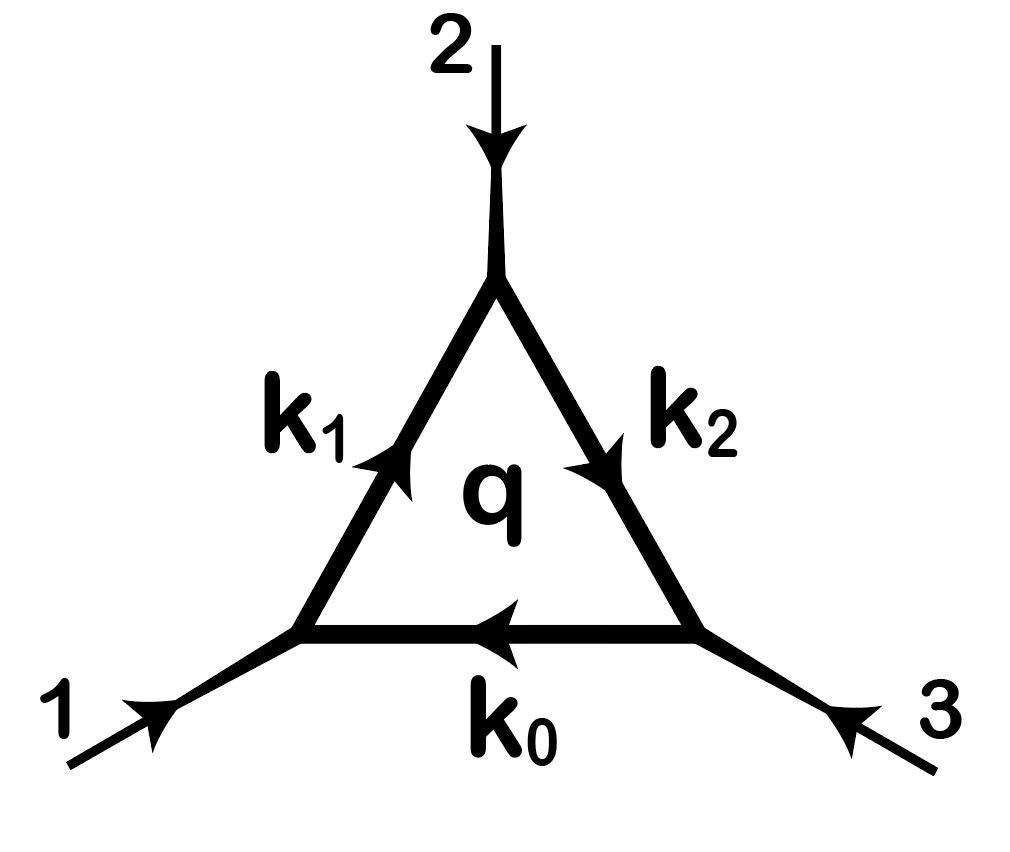}}\sim \nu_0\frac{(l_p)^{\Lambda_3-3}\PPb_{12}^{\Lambda_3}}{\prod_{i=1}^{3}\beta_i^{\lambda_i}\Gamma(\Lambda_3-2)\pvec_3^2}\,,
\end{align}
where $\pvec_3$ is the off-shell momentum and the expression is similar to the $\Gamma^{+++}$ amplitude for QCD \cite{Chakrabarti:2005ny}. We see that the overall factor $\nu_0$ makes it vanish.

%%%%%%%%%%%%%%%%%%%%%%%%%%%%%%%%%%%%%%%%%%%%%%%%%%%%%%%%%%
\subsection{General Loops}
%%%%%%%%%%%%%%%%%%%%%%%%%%%%%%%%%%%%%%%%%%%%%%%%%%%%%%%%%%
All loop diagrams can be shown to vanish if the total number of degrees of freedom is regularized to zero. Indeed, any $l$-loop $n$-point diagram can be represented as the union of elementary sunshine diagrams   
\begin{align}\notag
   \Gamma_{n}\sim \parbox{2.2cm}{\includegraphics[scale=0.12]{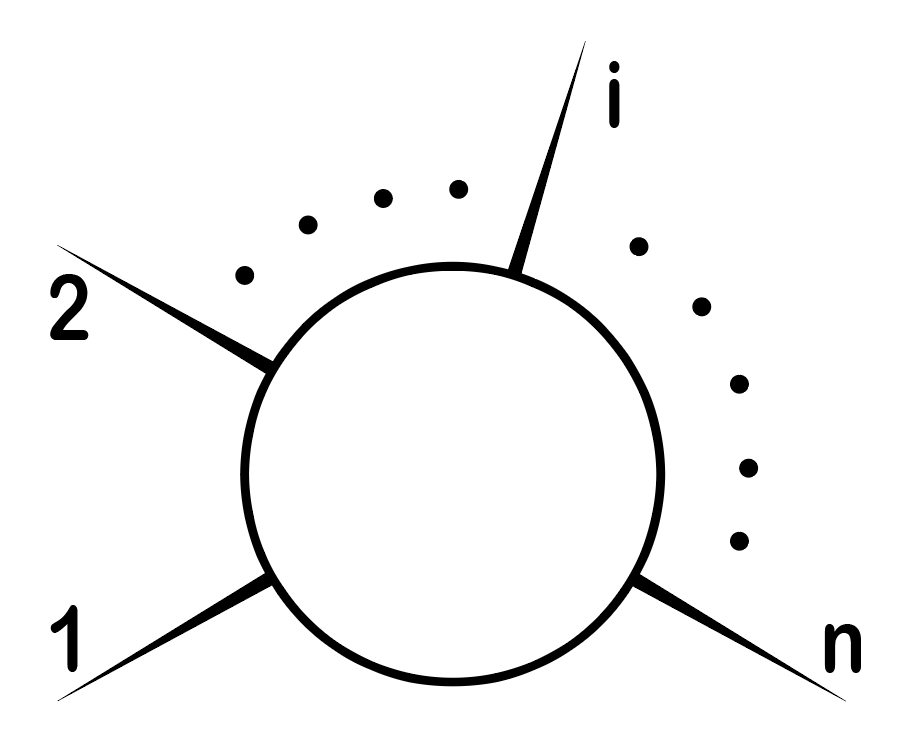}}
\end{align}
where the external legs may be off-shell and can be sewn with other sunshine diagrams. The lower order sunshine diagrams are the self-energy and the vertex correction given above. In general, in large-$N$ limit for $n>2$ we find an integral of the type 
\begin{equation}
    \Gamma_{n}\sim \nu_{0}\frac{1}{\prod_{i=1}^{n}\beta_i^{\lambda_i}}\int\frac{\mathcal{K}_n(\PPb_{ij})}{\pvec^2(\pvec+\pvec_1)^2 ... (\pvec-\pvec_{n})^2}
\end{equation}
with some numerator $\mathcal{K}_n$. What is important is that the 'total number of degrees of freedom' factors out in the form of $\nu_{0}$ \footnote{It would be interesting to see if the finite expressions for the loop diagrams (modulo the $\nu_0$-factor) can be given some interpretation.}. Therefore, each elementary sunshine diagram gets multiplied by zero if we assume \eqref{magic}, and hence any loop diagram also does so. As a result, all loop diagrams vanish, confirming that $S=1$. 

Let us note that the need to regularize the sums over fields is to be expected in any theory with infinitely many fields and represents one of the stringy features of the chiral \hsgra. Another instance of the same problem arises in dimensional reductions, see e.g. \cite{Fradkin:1982kf}, and Exceptional Field Theories \cite{Bossard:2015foa}. In general, it would be helpful to have some worldsheet realization of \hsgra's exempted from these regularization issues.

%%%%%%%%%%%%%%%%%%%%%%%%%%%%%%%%%%%%%%%%%%%%%%%%%%%%%%%%%%
\section{\label{sec:conc}Conclusions}
%%%%%%%%%%%%%%%%%%%%%%%%%%%%%%%%%%%%%%%%%%%%%%%%%%%%%%%%%%
We have studied the simplest chiral \hsgra{} with $u(N)$ Chan-Paton factors and showed that it is a fully consistent quantum \hsgra. Thanks to the coupling conspiracy, both the tree-level amplitudes and the loop corrections vanish. Therefore, $S=1$ to all orders in perturbation theory, which is consistent with the no-go theorems. This provides us with the first example of a quantum \hsgra. There are several variants of classical chiral \hsgra's \footnote{See \cite{Ponomarev:2017nrr}. We also expect SUSY chiral \hsgra{} to exist. Likewise, conformal chiral \hsgra{} should exist too.} and it would be interesting to see if all of them are quantum consistent. 

One may wonder if $S=1$ is a satisfactory answer or whether it means that \hsgra's are trivial. Firstly, this seems to be the only possible answer consistent with the no-go theorems. Secondly, pure \hsgra's are not meant to be realistic models of nature, rather they are toy models of quantum gravity whose importance is perhaps in having the minimal multiplet that allows the graviton to be embedded into a consistent quantum theory. More realistic models should result from matter-coupled and Higgsed \hsgra's, where the solution is expected to be string theory \cite{Caron-Huot:2016icg}. We argue below that the same reasoning should apply to AdS \hsgra's whenever they will be constructed and quantized.

Our findings for flat space can also shed some light on AdS \hsgra's. \hsgra{} in AdS \footnote{\hsgra{} in AdS is still at its infancy and few terms in the action are known \cite{Bekaert:2015tva, Sleight:2016dba} or only formally consistent equations are available \cite{Vasiliev:1990cm, Vasiliev:2003ev}. %The latter cannot represent a viable proposal even for a classical \hsgra{} due to locality not having been properly taken into account \cite{Boulanger:2015ova}. 
See \cite{Sharapov:2017yde,Bekaert:2017bpy,Grigoriev:2018wrx} for the recent progress in formal \hsgra.} are generic duals of free CFT's \cite{Sundborg:2000wp,Sezgin:2002rt,Klebanov:2002ja}. Indeed, gauge symmetries of massless higher spin fields in AdS translate into dual operators being conserved tensors. The charges associated with the latter signal an extension of the conformal symmetry. The AdS/CFT analog \cite{Maldacena:2011jn,Boulanger:2013zza,Alba:2013yda,Alba:2015upa} of the Coleman-Mandula theorem \cite{Coleman:1967ad} states that a CFT with a higher spin current is a free one in $d>2$. In any free CFT, say with a free field $\phi$, one can construct infinitely many higher spin currents $J_s$ as bilinears $J_s= \phi \partial...\partial \phi+...$. The fields of the dual \hsgra{} in AdS space are in one-to-one correspondence with $J_s$ and bulk interactions should  account for nonvanishing $\langle J...J\rangle$, which are built of free partons $\phi$. 

Therefore, being dual to a free CFT is a good generalization of the $S=1$ statement from flat space to AdS/CFT holographic $S$-matrix: asymptotic higher spin symmetries in flat space or in AdS imply $S=1$ or $S=\text{free CFT}$, respectively \footnote{Indeed, both statements are consequences of the linearized gauge symmetry $\delta \Phi_{\mu_1...\mu_s}=\nabla_{\mu_1}\xi_{\mu_2...\mu_s}+\text{\perm}$ at the infinity (conformal boundary) of flat (AdS) spaces.}. 
Based on the analogy above, our conjecture is that AdS \hsgra{} should have better UV behaviour (compared to the naive power counting) and the systematic reason for the loop corrections to vanish (or be proportional to the tree-level result) should be a factorization of sums over spins (one-loop bubbles), as occurs for the chiral \hsgra.

Lastly, there are non-\hsgra{} examples that share some of the properties of the chiral \hsgra: self-dual Yang-Mills and self-dual Gravity. They have vanishing tree level amplitudes and finite loop corrections, even though the reasons for cancellation seem to be somewhat different \cite{Chalmers:1996rq,Krasnov:2016emc}. 
Closer to the chiral \hsgra{} are the conformal \hsgra's \cite{Segal:2002gd,Tseytlin:2002gz}, which are defined in even dimensions as the local part of the induced action of a free CFT in the higher spin background. They should have vanishing tree-level amplitudes and give examples of consistent quantum conformal \hsgra{} \cite{Bekaert:2010ky,Joung:2015eny,Beccaria:2016syk}. See also \footnote{A somewhat similar example (vanishing tree-level amplitudes) was constructed in \cite{Fotopoulos:2007nm}. Tree level quantum properties of off-shell higher spin ``string-like'' systems  were considered in \cite{Fotopoulos:2010ay,Taronna:2011kt,Dempster:2012vw}. Certain quantum properties of hypothetical \hsgra{} in flat space were studied in \cite{Ponomarev:2016jqk}. See also  \cite{Roiban:2017iqg,Taronna:2017wbx}.}.

\begin{acknowledgments}
We are grateful to Sudarshan Ananth, Nicolas Boulanger, Andrea Campoleoni, Dario Francia, Kirill Krasnov, Ruslan Metsaev, Radu Roiban, Arkady Tseytlin and especially to Dmitry Ponomarev for very useful discussions. The work of E.S. was supported by the Russian Science Foundation grant 14-42-00047 in association with the Lebedev Physical Institute. The work of T.T. was supported by the DFG Transregional Collaborative Research Centre TRR 33 and the DFG cluster of excellence ``Origin and Structure of the Universe". The work of MT was supported by the ARC grant DP160103633.
\end{acknowledgments}
\appendix

%%%%%%%%%%%%%%%%%%%%%%%%%%%%%%%%%%%%%%%%%%%%%%%%%%%%%%%%%%
\section{\label{app:ident}Identities}
%%%%%%%%%%%%%%%%%%%%%%%%%%%%%%%%%%%%%%%%%%%%%%%%%%%%%%%%%%
The light-cone kinematical variables obey a number of identities \cite{Chakrabarti:2005ny} that are indispensable for the computations in the paper. Firstly, there Bianchi-like identities (valid for $\PP$ and $\PPb$):
\begin{align}\notag
    \sum_i \PP_i&=0\,,&
    \beta_{[i} \PP_{jk]}&\equiv0\,,&
    \PP_{i[j}\PP_{kl]}&\equiv0\,.
\end{align}
Other useful kinematic identities include
\begin{align}\notag
    \sum_j \frac{\PP_{ij} \PPb_{jk}}{\beta_j}&= -\frac12\beta_i\beta_k \sum_j \frac{\pvec_j^2}{\beta_j}\,,
\end{align}
and for $\pvec_i^2=0$, $\pvec_k^2\neq 0$ we have
\begin{equation}\notag
    \PPb_{ik}\PP_{ik}=-\frac{\beta_i\beta_k}{2}(\pvec_i+\pvec_k)^2+\frac{1}{2}\beta_i(\beta_k+\beta_i)\pvec_k^2\,.
\end{equation}

\end{document}